\documentclass{aastex631}

\shorttitle{The quantum basis of global warming}
\shortauthors{Wordsworth et al.}

\usepackage{graphicx}
\usepackage{dcolumn}
\usepackage{bm}
\usepackage[version=3]{mhchem} 
\def\D{\mathrm{d}}
\def\E{\mathrm{e}}
\def\Wmsq{W~m$^{-2}$}
\def\invcm{cm$^{-1}$}
\def\Wm2K{W~m$^{-2}$~K$^{-1}$}
\def\mO{m_{\ce{O}} }
\def\mC{m_{\ce{C}} }
\def\Nm{N~m$^{-1}$ }

\begin{document}

\title{Fermi Resonance and the Quantum Mechanical Basis of Global Warming}

\author{R.~Wordsworth}
\affiliation{School of Engineering and Applied Sciences, Harvard, Cambridge, MA 02138, USA}
\affiliation{Department of Earth and Planetary Sciences, Harvard, Cambridge, MA 02138, USA}
\email{rwordsworth@seas.harvard.edu}

\author{J.~T.~Seeley}
\affiliation{School of Engineering and Applied Sciences, Harvard, Cambridge, MA 02138, USA}

\author{K.~P.~Shine}
\affiliation{Department of Meteorology, University of Reading, Reading, RG6 6ET, United Kingdom}

\begin{abstract}
Although the scientific principles of anthropogenic climate change are well-established, existing calculations of the warming effect of carbon dioxide rely on spectral absorption databases, which obscures the physical foundations of the climate problem. Here we show how \ce{CO2} radiative forcing can be expressed via a first-principles description of the molecule's key vibrational-rotational transitions. Our analysis elucidates the dependence of carbon dioxide's effectiveness as a greenhouse gas on the Fermi resonance between the symmetric stretch mode $\nu_1$ and bending mode $\nu_2$. It is remarkable that an apparently accidental quantum resonance in an otherwise ordinary three-atom molecule has had such a large impact on our planet's climate over geologic time, and will also help determine its future warming due to human activity. {In addition to providing a simple explanation of \ce{CO2} radiative forcing on Earth, our results may have implications for understanding radiation and climate on other planets.}
\end{abstract}


\section{\label{sec:intro}Introduction}

{Carbon dioxide is an essential greenhouse gas on all rocky planets in the solar system with significant atmospheres (Venus, Earth and Mars). On Earth, the carbonate-silicate cycle regulates atmospheric \ce{CO2} on geological timescales, but}  the last 150 years has seen a rapid rise in concentrations from approximately $280$~ppmv to $415$~ppmv \citep{tziperman2022global},  
due to burning of fossil fuels by humans and land use changes \citep{friedlingstein2022global}.  
Earth's global mean surface temperature has risen by approximately 1~K during this same period, with most of the warming a direct result of this \ce{CO2} increase. \ce{CO2} affects surface temperature because it is a greenhouse gas: it absorbs more {effectively} at thermal infrared frequencies than the near-infrared and visible frequencies where solar radiation peaks. As a result, increasing levels of atmospheric \ce{CO2} shifts the emission of thermal radiation to space to higher altitude regions of the atmosphere, where air is less dense and colder. This colder air releases less thermal radiation, so increasing \ce{CO2} decreases total emission to space for fixed surface and atmospheric temperatures. The magnitude of this decrease is defined as the radiative forcing of \ce{CO2}.

\begin{figure}[h!]
	\begin{center}
    \includegraphics[width=2.5in]{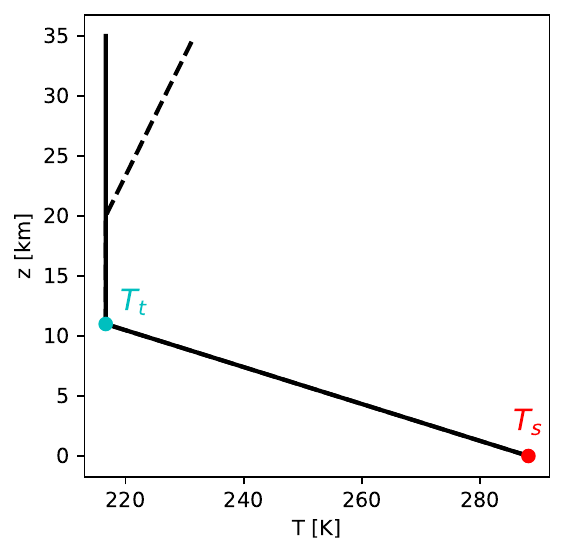}
	\end{center}
	\caption{Idealized plot of temperature vs. altitude in Earth's atmosphere. The solid black line shows the temperature structure used to derive \eqref{eq:alpha_defn} {(fixed tropospheric lapse rate of -6.5 K~km$^{1}$, isothermal stratospheric temperature of 217~K)}. The dashed black line shows the 1976 Standard Atmosphere temperature profile \citep{minzner19771976}. Red and cyan dots show the values of $T_s$ and $T_t$, respectively.}
\label{fig:atm_T_prof}
\end{figure}

\begin{figure}[h!]
	\begin{center}
	\includegraphics[width=6.5in]{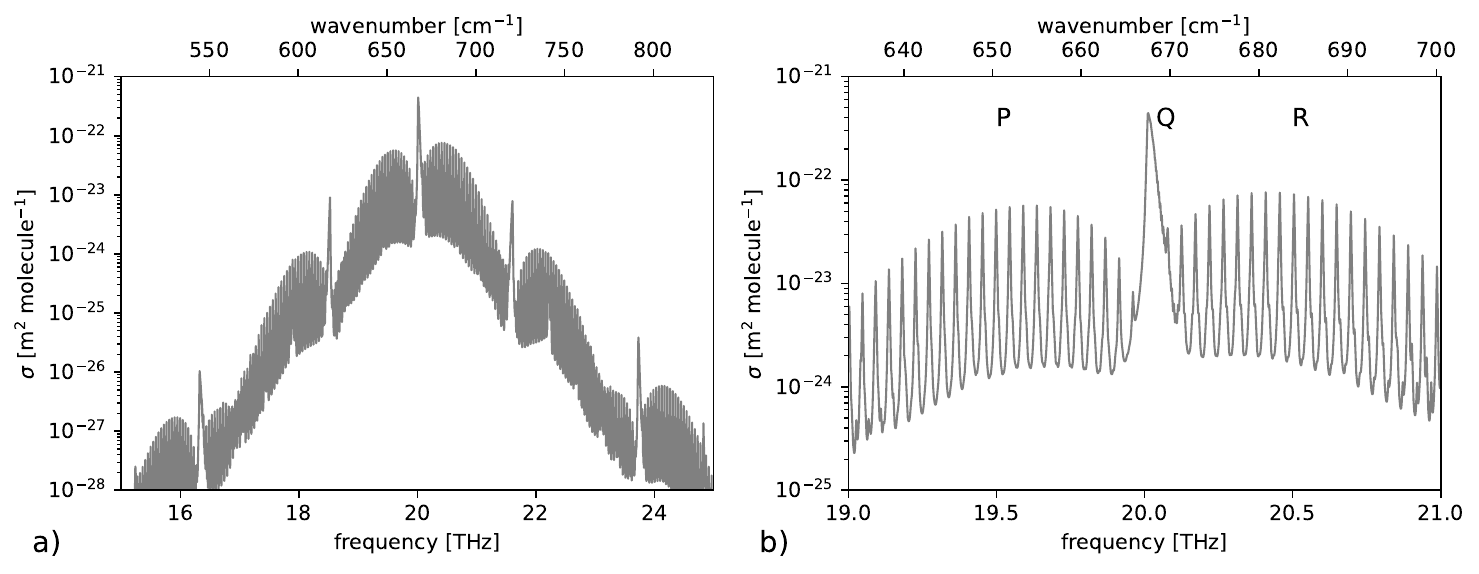}
	\end{center}
	\caption{High-accuracy absorption spectrum of \ce{CO2} in the region of the $\nu_2$ band, in cross-section units of meters squared per molecule of \ce{CO2}, at pressure $p_s$ and temperature $T_s$ (see Table~1). a) shows the band over a wide frequency range, while b) zooms in on the central fundamental band, with P, Q and R branches labeled (see Section~\ref{sec:method}). Both plots were produced using the HITRAN database, with line truncation at the standard value of 25~\invcm (0.75~THz) \citep{gordon2022hitran2020}. }
\label{fig:HITRAN_spec}
\end{figure}

Atmospheric mixing is fast compared to the rate of \ce{CO2} emission and removal, so to a first approximation the \ce{CO2} concentration is uniform in Earth's lower atmosphere. The stratosphere-adjusted radiative forcing\footnote{{This differs slightly from the \emph{effective} radiative forcing definition now used by the IPCC \citep{masson2021climate}, although the difference is not important for our purposes. See also discussion in Section~\ref{sec:radforce}.}} due to an increase in the atmospheric \ce{CO2} molar concentration from $x_0$ to $x$ is typically expressed as 
\begin{equation}
\Delta F \approx +\alpha \ln [x/x_0] \label{eq:DeltaF}
\end{equation}
where $\alpha$ is defined as the \ce{CO2} radiative forcing parameter. Global-mean calculations from {detailed} radiative transfer codes using tabulated spectroscopic data yield $\alpha \approx 5.35$~\Wmsq~\citep{myhre1998new}, {with an estimated accuracy of around 10\%}. {Once the atmosphere and ocean have thermally equilibrated}, this radiative forcing gives rise to a surface temperature change of magnitude 
\begin{equation}
\Delta T_s \approx +\lambda^{-1} \Delta F \label{eq:ECS}
\end{equation}
where the parameter $\lambda$ is the climate feedback parameter, with units of \Wm2K. Based on data from a range of observational and modeling sources, the sixth assessment report of the Intergovernmental Panel on Climate Change (IPCC AR6) stated that $\Delta T_s$ for a doubling of \ce{CO2} is``very likely" (i.e., with greater than 90\%  probability) in the range 2 to 5~K, with a best estimate of 3 K \citep{masson2021climate}.  Since the radiative forcing for \ce{CO2} doubling $\Delta F_{2\times} = 5.35 \ln(2) = 3.71$~\Wmsq, this indicates a value of between 0.74 and 1.85~\Wm2K for $\lambda$.

Given the clear correspondence between observations and the results of sophisticated climate models, the scientific basis of climate change is indisputable. In addition, many comprehensive descriptions of the physics of climate and global warming, including the specifics of the radiative effects of \ce{CO2} doubling, already exist \citep{pierrehumbert2011principles,wilson2012simple,zhong2013greenhouse,mlynczak2016spectroscopic,dufresne2020greenhouse,jeevanjee2021analytical,romps2022forcing,tziperman2022global,shine2023radiative}. Despite this, it is currently still not possible to derive \eqref{eq:DeltaF} directly starting from fundamental properties of the \ce{CO2} molecule. This is an important objective, because analytic methods are a powerful tool to increase understanding and elucidate the results of numerical simulations. Here, we build on previous efforts and show how this can be achieved, via a synthesis of molecular spectroscopy and climate physics. {Our analysis here focuses on Earth's present-day climate, but potential applications to other planets in the solar system and exoplanets are discussed in Section~\ref{sec:climsens}}.

\section{\label{sec:radforce}A simple empirical model of \ce{CO2} radiative forcing}

In several recent papers \citep{wilson2012simple,jeevanjee2021analytical,romps2022forcing}, it has been shown that key features of \ce{CO2} radiative forcing can be captured using a simplified representation of atmospheric radiative transfer combined with an empirical approach to \ce{CO2} spectroscopy. Absorption by \ce{CO2} in the thermal infrared at Earth-like concentrations is dominated by a broad collection of bands centered around 20~THz (667~cm$^{-1}$ in wavenumber units)\footnote{Traditionally, one of the novelties of atmospheric spectroscopy work is a continuous need to interconvert between various non-SI units. Here, we stick to SI units as much as possible, but when stating frequencies in Hz or THz we also report wavenumber values in cm$^{-1}$, to allow easy comparison with other work.} that has a `triangular' shape in logarithmic units (Fig.~\ref{fig:HITRAN_spec}). Because of this, this collection of bands can approximately be represented using the formula
\begin{equation}
\ln[\sigma/\sigma_{cen}]=-\frac{|\nu-\nu_{cen}|}w\label{eq:band_struc_init}
\end{equation}
for absorption cross-section $\sigma$. Here $\sigma_{cen}$ is the value of $\sigma$ at the band center, $\nu$ is frequency, $\nu_{cen}$ is the frequency of the band center, and $w$ is a band structure coefficient, such that a smaller value of $w$ leads to a narrower band. Numerical estimation of  $w$ from HITRAN spectroscopic data {or line-by-line spectral radiative forcing calculations} yields a value of around 0.37~THz (12.5~cm$^{-1}$), with the value somewhat dependent on the fitting approach chosen \citep{jeevanjee2021analytical,romps2022forcing}.

Taking the angle-averaged optical depth $\tau$ as increasing downwards from the top of the atmosphere to the surface, we can write a small change in $\tau$ as 
\begin{equation}
d\tau = \frac{ x \sigma  }{  \overline \mu  m_a }\frac{dp}g
\end{equation}
where $\overline \mu$ is the mean propagation angle of upwelling infrared photons, $p$ is pressure, $g$ gravity and $m_a$ is the mean molecular mass of air. If we assume a linear dependence of $\sigma$  on pressure due to line broadening\footnote{This is a simplified approach, because pressure scaling at line centers and line wings differs \cite[see also Section~\ref{sec:linewidth}]{goody1995atmospheric}. However, the approach works well in practice \citep{romps2022forcing}, and the specific choice of pressure scaling in this equation does not affect our subsequent analysis.}  such that $\sigma = \sigma_0(p/p_0)$, where $p_0$ is a reference pressure (Table~1), then
\begin{equation}
\tau = \frac{x \sigma_0 p^2}{2g \overline {\mu} m_a p_0}. \label{eq:opt_depth}
\end{equation}
Next, approximating infrared emission to space at a given frequency as coming from a narrow pressure range in the atmosphere, one can write an emission pressure corresponding to $\tau(p,\nu) = \tau_{em}(\nu)$. Combining \eqref{eq:band_struc_init} and \eqref{eq:opt_depth},
{noting that $\tau$ and $\sigma$ are proportional,} setting $\tau_{em} = 1$, rearranging in terms of $\nu$ and writing $\nu = \nu_{em}$ yields \citep{jeevanjee2021analytical}
\begin{equation}
\nu_{em}(p,x) =\nu_{cen} \pm w \ln\left[\frac{x \sigma_{cen,0} p^2}{2g \overline {\mu} m_a p_0}\right].
\end{equation}
The logarithmic dependence of $\nu_{em}$ on $x$ is what gives rise to the logarithmic dependence of $\Delta F$ on $x$. The relationship between $\nu_{em}$ and $\Delta F$ can be determined by noting that increasing $x$ is equivalent to swapping emission over a certain frequency range from the surface to the stratosphere \citep{romps2022forcing}. The expression for the \ce{CO2} radiative forcing parameter that emerges from this analysis is
\begin{equation}
\alpha =  2\pi w[ \mathcal  B(\nu_{cen},T_s) -  \mathcal  B(\nu_{cen},T_t)]. \label{eq:alpha_defn}
\end{equation}
Here $\mathcal  B(\nu,T)$ is the Planck spectral irradiance evaluated at frequency $\nu$ and temperature $T$, $T_s$ is surface temperature and $T_t$ is the tropopause temperature (see Fig.~\ref{fig:atm_T_prof}).

The analytic model represented by \eqref{eq:alpha_defn} is highly simplified, but it allows greater insight into the mechanism of \ce{CO2} radiative forcing than is possible from  numerical approaches. Like all current approaches to understanding climate change, however, it still requires us to assume that the infrared spectrum of \ce{CO2} is available as a pre-determined input. Our aim here is to relax this requirement, and {derive} $w$ and $\nu_{cen}$, and hence equations \eqref{eq:DeltaF} and \eqref{eq:alpha_defn}, {from} the basic properties of the \ce{CO2} molecule.

\section{\label{sec:method}\ce{CO2} infrared spectroscopy}

\begin{figure}[h!]
	\begin{center}
	\includegraphics[width=6.5in]{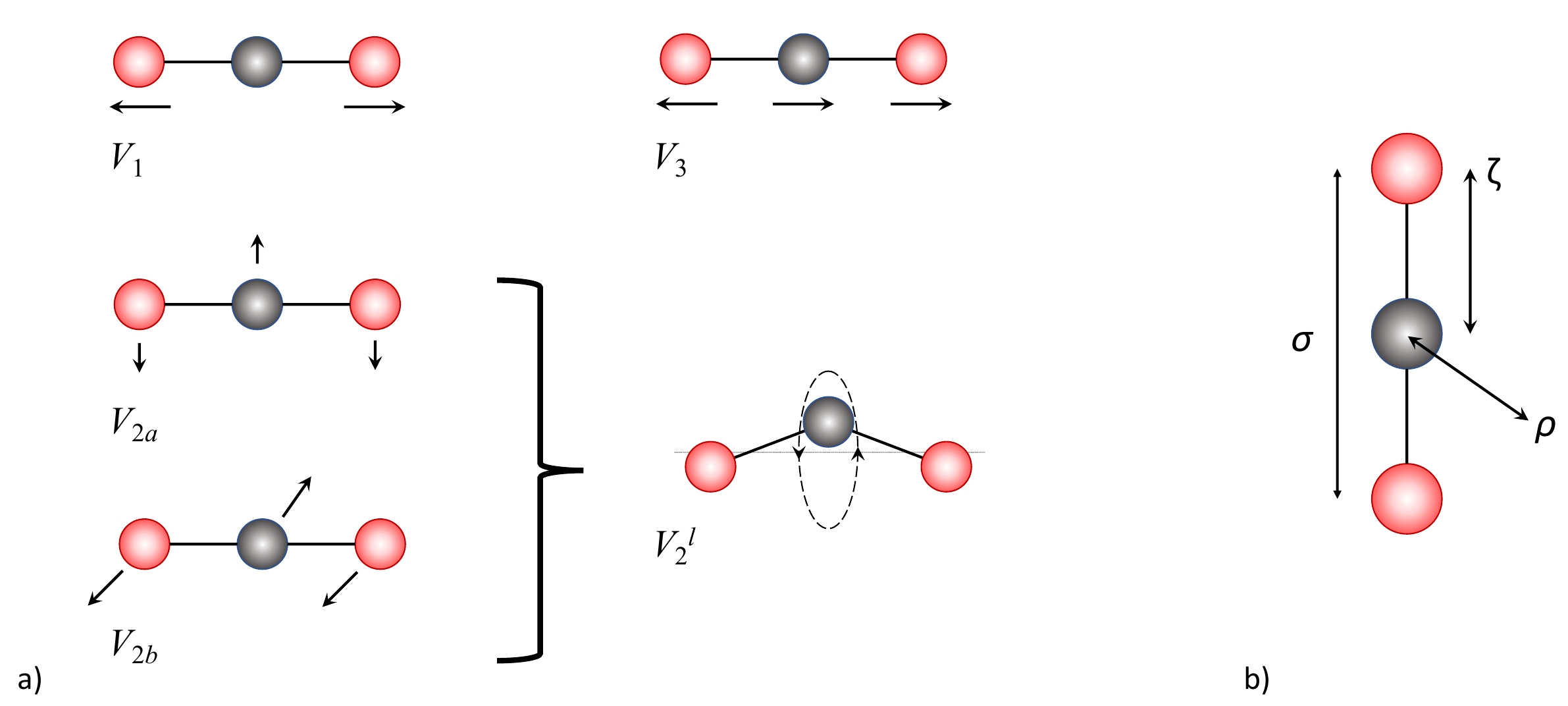}
	\end{center}
	\caption{a) Schematic of the three vibrational modes of carbon dioxide. The two degenerate bending modes superimpose to produce a motion where each atom rotates around the major axis of the molecule, which is represented via the quantum number $l$. b) Mass-weighted coordinate system used to express the three fundamental modes of \ce{CO2} as simple harmonic oscillations (see Section~\ref{sec:Fermi}).}
\label{fig:mode_schematic}
\end{figure}

\ce{CO2} absorbs in the infrared due to combinations of vibrational and rotational quantum transitions \citep{pierrehumbert2011principles}. As a three-atom linear molecule, \ce{CO2} has $3N-5 = 4$ vibrational degrees of freedom (Fig.~\ref{fig:mode_schematic}), with four quantum numbers $V_1$, $V_{2a}$, $V_{2b}$, and $V_3$ corresponding to excitation of an symmetric stretch mode, two degenerate bending modes, and an asymmetric stretch mode, respectively\footnote{The use of $\nu$ as the symbol for frequency and $v$ for vibrational quantum number is common in the modern spectroscopic literature, but to avoid confusion we have chosen to use the upper-case $V$ notation of \cite{adel1933infrared} for quantum number here.}. Superposition of the in-plane and out-of-plane bending motions corresponding to $V_{2a}$ and $V_{2b}$ results in an excitation where the three atoms perform circular motions about the molecule's major axis (Fig.~\ref{fig:mode_schematic}). Such motion has angular momentum, which can be represented via introduction of a new quantum number $l$. Because $V_{2a}$ and $V_{2b}$ are degenerate, vibrational \ce{CO2} states can therefore be characterized\footnote{As always, the Dirac or bra-ket notation represents quantum states, such that $|\psi \rangle = \int d^3\mathbf x \psi(\mathbf x) |\mathbf x \rangle$, and $\langle \psi |\phi \rangle = \int  \psi(\mathbf x)^* \phi(\mathbf x) d^3\mathbf x$ when written in terms of position $\mathbf x$ and wavefunctions $\psi$ and $\phi$.} as $|V_1V_2^lV_3\rangle$, where $V_2 = V_{2a} + V_{2b}$ and $l = V_2, V_2 - 2,...,1$ or 0. Hence for $V_2 = 3$, for example, $l$ can be 3 or 1, while for $V_2=4$, $l=4$, 2 or 0 are all allowed.

The frequency of each vibrational mode of \ce{CO2} can be calculated approximately\footnote{More accurate determination of vibrational frequencies can be achieved via \emph{ab initio} methods \citep[e.g., ][]{rodriguez2007fermi}.} by treating the molecule as three point masses connected by spring-like bonds. In particular, using a valence force approach, which assumes restoring forces act to oppose changes in the distance and angles of local atom-atom bonds only \citep[e.g., ][p. 168]{herzberg1945molecular}, the molecular vibrational potential energy $U$ can be written as  
\begin{equation}
U = U_e + \frac 12 k (Q_1^2+Q_2^2) + \frac 12 k_\delta \delta^2.
\label{eq:U}
\end{equation}
Here $U_e$ is the equilibrium potential energy corresponding to zero vibrational motion, while $Q_1$ and $Q_2$ are the change in separation between the C and the first and second O atoms during vibrational motion, respectively. $\delta$ is the change in bond angle that occurs during bending motion, and $k$ and $k_\delta$ are linear and bending bond force constants, respectively. 

The form of the terms in \eqref{eq:U} allow the normal mode frequencies of the molecule to be simply expressed. We focus on the first two modes here. While the antisymmetric stretch mode $\nu_3$ gives rise to a strong absorption band, it is located outside the thermal infrared spectral region and hence is not relevant to our discussion. Additional combination bands involving $\nu_3$ are present around 30~THz (1000~\invcm) but only contribute around 5\% to radiative forcing at present-day \ce{CO2} concentrations \citep{mlynczak2016spectroscopic}. 

For the symmetric stretch mode, we simply have oscillatory motion of each O atom, leading to 
\begin{equation}
\nu_1 = \frac 1{2\pi}\sqrt{\frac k \mO} 
\end{equation}
where $\mO$ is the mass of an oxygen atom. The value {$k=1680$}~\Nm yields $\nu_1 = 40.1$~THz (1337~\invcm). This is fairly close to the C-O bond force constant of 1860~\Nm in carbon monoxide (CO). 
 The bending mode frequency is 
\begin{equation}
\nu_2 = \frac{1}{2\pi}\sqrt{\frac 2 \mO \left(1 + 2 \frac{m_{\ce{O}}}{m_{\ce{C}}}\right) \frac{k_\delta}{a_e^2}} 
\label{eq:nu2_analytic}
\end{equation}
where $\mC$ is the mass of a carbon atom \citep{herzberg1945molecular}. For $k_\delta=7.7\times10^{-19}$~N~m and $a_e=116$~pm (Table~1), 
the value of $\nu_2$ is 20.0~THz (667~cm$^{-1}$). $\nu_2$ is of course the well-known bending mode associated with the center of the thermal infrared band, $\nu_{cen}$, as seen in Fig.~\ref{fig:HITRAN_spec}. The fact that $\nu_2$ is (apparently coincidentally) about half the value of $\nu_1$ will be very important in the discussion that follows. 

\begin{table}[h]
\centering
\begin{tabular}{cccc}
\hline
\hline
Parameter & Symbol & Value & Units   \\
\hline
Mass of oxygen atom & $m_{\ce O}$ & 16 & $m_u$    \\
Mass of carbon atom & $m_{\ce C}$ & 12 & $m_u$    \\
Equilibrium C--O separation in \ce{CO2} & $a_{e}$ & $1.16\times10^{-10}$ & m    \\ 
\ce{CO2} symmetric stretch force constant & k & $1680$ & N~m$^{-1}$    \\ 
\ce{CO2} bending mode force constant & $k_\delta/a_e^2$ & $57$ & N~m$^{-1}$     \\ 
\ce{CO2} transition dipole moment magnitude & $|\langle m|\mathbf d| n \rangle|$ & $3.35\times10^{-31}$ & C~m \\ 
\ce{CO2} Fermi coupling term & $|b|$ & $2.14\times10^{12}$ & Hz \\ 
Collision cross-section (\ce{CO2} in \ce{N2}) & $\sigma_c$ & $0.44\times10^{-18}$ & m$^2$ \\ 
Line broadening coefficient & $n_b$ & 0.5 & []    \\ 
Earth mean surface pressure & $p_s$ & $10^5$ & Pa    \\
Earth mean surface temperature & $T_s$ & 288 & K    \\
Earth stratospheric temperature & $T_t$ & 217 & K    \\
\hline
\hline
\end{tabular}
\newline
\newline
Table 1:~Key parameters used in the analysis. Throughout this paper, we assume that reference pressure and temperature are equal to their surface values, $p_0 \equiv p_s$ and $T_0 \equiv T_s$. $m_u$ is the atomic mass unit such that 1~$m_u = 1.66\times10^{-27}$~kg. Values for $a_e$, $k$, $k_\delta/a_e^2$ and $|b|$ are taken from \cite{herzberg1945molecular}, pp. 21, 173 \& 218; $\sigma_c$ is from \cite{chapman1990mathematical}, p. 263. Values for other quantities are justified in the main text.
\label{tab:params}
\end{table}

\subsection{Line positions} \label{sec:linepos}

Now we have an expression for the spectral location of the $\nu_2$ band in Fig.~\ref{fig:HITRAN_spec}, we can start to build up a description of the band structure. For this, we need to know what determines the location, shape and intensity of all the most important spectral lines. We tackle this problem step by step, starting with line location. We focus on explaining the most abundant isotopologue of \ce{CO2}, \ce{^{12}C^{16}O2},  on the basis that the contribution of all other isotopologues to \ce{CO2} radiative forcing is minor \citep{shine2023radiative}.

In the infrared, spectral lines appear because absorption or emission of photons causes quantized changes in the vibrational and/or rotational state of molecules. The center of the $\nu_2$ band\footnote{We use the term `$\nu_2$ band' here and elsewhere for simplicity, but as will become clear soon, transitions involving additional vibrational modes are also important to the larger band structure.} at 20~THz (667~\invcm) corresponds to a vibrational transition from the ground state $|00^00\rangle$ to the first excited bending mode with angular momentum number $l=1$, $|01^10\rangle$. The band has a strong central peak called the Q-branch (Fig.~\ref{fig:HITRAN_spec}) due to purely vibrational transitions, and P and R branches at lower and higher frequencies, respectively, due to combinations of vibrational and rotational transitions. Solution of the time-independent Schr\"odinger equation in spherical coordinates \citep{levine1975molecular} shows that the location of P- and R-branch lines relative to $\nu_2$ is given by 
\begin{equation}
\nu_{J,J+1} =  (E_{J+1}-E_J)/h = 2B(J+1).
\end{equation}
Here $h$ is Planck's constant, $J=0,1,2,...$ is the rotational quantum number, $E_J$ is the energy of rotational state $J$, and the rotational constant $B=h/8 \pi^2 I$, 
with moment of inertia $I=2m_{\ce{O}}a_e^2$.  {In addition, $a_e$ is the equilibrium C--O separation in \ce{CO2} (Table~1).}
For lines in the Q branch, there is no change in $J$, while in the P and R branches, $\Delta J = \pm 1$. For \ce{^{12}C^{16}O2}, transitions involving odd values of $J$ are missing, because of selection rules arising from the zero spin of the oxygen atoms \citep{levine1975molecular}. {Given that $a_e=116$~pm, $B=  h/16 \pi^2 m_{\ce{O}}a_e^2  =11.7$~GHz (0.39~\invcm)}, 
so the spacing between the P- and R-branch lines is about $2\times 2B=47$~GHz (1.6~\invcm).  In the Q-branch, transitions with different rotational energies are not exactly colocated because of subtle effects such as Coriolis interactions, but for our purposes the lines can be treated as {unseparated in frequency}. Rotational line spacing is further affected by centrifugal and anharmonic effects, but these complications are not important to climate forcing and so will be ignored here. 

\subsection{Line shape and width} \label{sec:linewidth}

All spectral lines have a shape that is determined by a combination of natural broadening, Doppler and collisional effects \citep{goody1995atmospheric}. Line shape is dependent on both temperature and pressure. A full analysis of this problem could become complicated quickly, but in keeping with our aim of getting a rough estimate of \ce{CO2} radiative forcing only, we take a simple approach here. In the troposphere, thermal infrared absorption lines are Lorentzian to a close approximation, with lineshape
\begin{equation}
f(\nu-\nu_{mn}) =\frac 1 \pi \frac{\gamma}{(\nu-\nu_{mn})^2 + \gamma^2 }
\end{equation}
where $\nu_{mn}$ is the frequency of the transition from state $m$ to $n$ and $\gamma$ is the linewidth (halfwidth at half maximum) in Hz. $\gamma$  scales approximately linearly with pressure. This can be shown by noting that the root mean square speed of a molecule in a gas of temperature $T$ is 
\begin{equation}
\overline v = \sqrt{3k_B T/\overline m}
\end{equation}
where $k_B$ is Boltzmann's constant, and $\overline m$ is the mean molecular mass of air. The mean relative collision speed is larger than this by a factor of $\sqrt{2}$, $\overline v_{rel} = \sqrt{2} \overline v$. The mean free path, or average distance travelled by a molecule between collisions, is \citep[p. 88]{chapman1990mathematical}
\begin{equation}
l_{mfp} = \frac{1}{\sqrt{2}n \sigma_c}= \frac{k_B T}{\sqrt{2} \sigma_c p}
\end{equation}
where $\sigma_c$ is intermolecular collision cross-section, $n$ is number density, $T$ is temperature, $p$ is pressure, and $p=nk_BT$ from the ideal gas law. The linewidth in frequency units can be written as $\gamma = 1/(2\pi\tau_c) = \overline v_{rel} / (2\pi l_{mfp})$, {where $\tau_c$ is the mean collision time between molecules}\footnote{{The factor of $2\pi$ in the relationship between $\gamma$ and $\tau_c$ can be obtained from a Fourier analysis of an ensemble of radiating oscillators \citep[e.g., ][p. 88]{stamnes2017radiative}}.}. 
Finally we have 
\begin{equation}
\gamma(p,T) = \gamma_0 (p/p_0) (T/T_0)^{-n_b}.\label{eq:gamma_eqn}
\end{equation}
Here $n_b = 0.5$, the reference values $p_0$, $T_0$ are taken to be surface values (Table~1), and 
\begin{equation}
\gamma_0 = \frac{\sigma_c p_0}\pi \sqrt{\frac 3 {\overline m k_B T_0}}. \label{eq:gamma0_eqn}
\end{equation}
For \ce{CO2} in \ce{N2}, $\sigma_c=\pi \times (3.75\times10^{-10}\mbox{ m})^2=0.44$~(nm)$^2$ \citep{chapman1990mathematical}, so
$\gamma_0=1.76$~GHz ($0.06$~\invcm). The HITRAN database \citep{gordon2017hitran2016} gives values for $\gamma_0$ in the \ce{CO2} $\nu_2$ band that range between 1.5 and 3~GHz (0.05~\invcm and 0.1~\invcm), and $n_b$ between 0.5 and 0.8, so our simple method slightly underestimates $\gamma_0$ and $n_b$. These differences are not important for the analysis that follows.

\subsection{Line intensity} \label{ssec:lineintensity}

Determining line intensity is one of the most challenging aspects of quantum spectroscopy. However, in the range of \ce{CO2} concentrations over which radiative forcing scales approximately logarithmically according to \eqref{eq:DeltaF}, $\Delta F$ is not sensitive to the absolute values of line intensity in the $\nu_2$ band \citep[e.g., ][]{jeevanjee2021analytical}. Hence we can also take an approximate approach here. The most important quantities in line intensity calculations are the Einstein coefficients, which express the rate of absorption or emission of a photon by a \ce{CO2} molecule. Analysis of the Schr\"odinger equation in the presence of a time-dependent perturbation due to an oscillating electric field leads to the following expression for the Einstein coefficient for spontaneous emission\footnote{In many textbooks, 
the Einstein-A coefficient is written in cgs units as $A_{mn}=64\pi^4 \nu_{mn}^3 |\langle m |\mathbf d | n \rangle|^2 / 3 h c^3$. To convert, note that in cgs units $\epsilon_0 = 1/4\pi$, which when substituted into the previous expression yields \eqref{eq:EinsteinA}. 
}:
\begin{equation}
A_{mn} = \frac{16\pi^3\nu_{mn}^3|\langle m |\mathbf d | n \rangle|^2}{3 h c^3 \epsilon_0}. \label{eq:EinsteinA}
\end{equation}
Here $m$ and $n$ are any two quantum states of energy $E_m$ and $E_n$ such that $E_m>E_n$ and $\nu_{mn} = (E_m-E_n)/h$. In addition, $c$ is the speed of light and $\epsilon_0$ is the vacuum permittivity. The term $\langle m |\mathbf d | n \rangle$ is the \emph{transition dipole moment}, which is defined as 
\begin{equation}
\langle m |\mathbf d | n \rangle \equiv \langle n |\mathbf d | m \rangle = \int \psi_m(\mathbf x)^*\mathbf d \psi_n(\mathbf x) d^3\mathbf x.
\end{equation}
Here $\mathbf d$ is the dipole moment {operator} of the molecule and $\psi_k$ is the eigenfunction of quantum state $k$. The transition dipole moment has typical magnitude $3.34\times10^{-31}$~C m ($0.1$~D) for \ce{CO2} transitions in the $\nu_2$ vibration-rotation band. This is similar to the permanent dipole of the ground state of CO, $4.07\times10^{-31}$~C.m (0.122~D) \citep{muenter1975electric}. Taking $\nu_{mn} = 20$~THz yields $A_{mn} \approx 1$~s$^{-1}$. Comparison with HITRAN data shows that this is a reasonable approximation at the center of the $\nu_2$ band, although the values of $A_{mn}$ decrease away from the band center.

\begin{figure}[b]
	\begin{center}
	\includegraphics[width=6.5in]{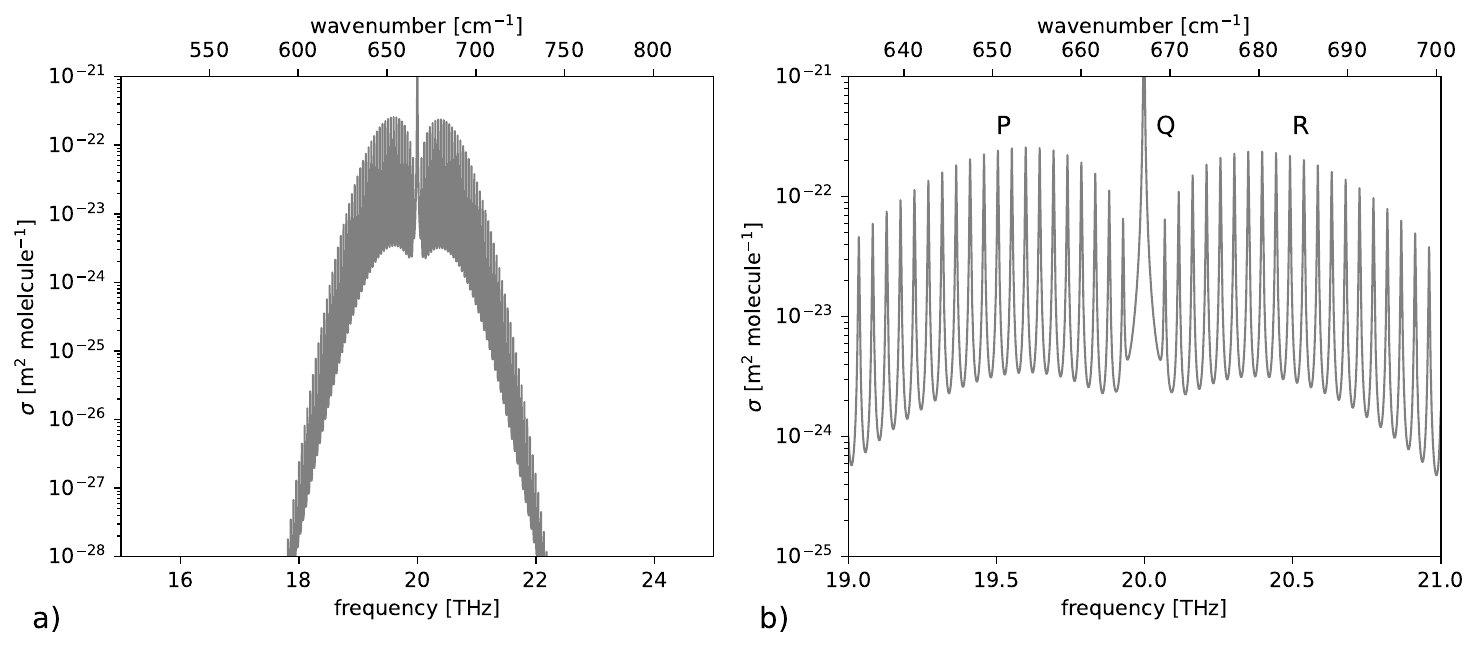}
	\end{center}
	\caption{a) Analytic \ce{CO2} $\nu_2$ band, including P, Q and R branches for the fundamental vibrational transition from $|00^00\rangle$ to $|01^10\rangle$, but neglecting all additional vibrational transitions. b) shows a close-up of the same band. Pressure and temperature are $p_s$ and $T_s$ in Table~1, as in Fig.~\ref{fig:HITRAN_spec}.}
\label{fig:analytic_spec}
\end{figure}

Line intensity expresses the frequency-integrated absorption cross-section of a line, independent of lineshape. In SI units of Hz~m$^2$~molecule$^{-1}$, it is defined as\footnote{In HITRAN notation \citep{simeckova2006einstein}, we would write $\tilde S_{mn} = ({A_{mn}}/{8\pi c\tilde \nu_{mn}^2})\mathcal O(T)$, where $\tilde S_{mn}$ has units of cm$^{-1}$~(cm$^2$~molecule$^{-2}$). To convert, we use $\tilde \nu = \nu/ c$, and in addition  $ S_{mn} = c\tilde S_{mn}$.}
\begin{equation}
S_{mn} = \frac{c^2A_{mn}}{8\pi \nu_{mn}^2}\mathcal O_{mn}(T).
\end{equation}
The dimensionless term $\mathcal O_{mn}(T)$ incorporates multiple additional effects, of which the most important here is energy level occupancy. The rate of absorption of photons by molecules in a given energy state must depend on the number density of molecules in that state, relative to the total number density. At temperatures in the 250-290~K range, most \ce{CO2} molecules are in the $\nu_2$ vibrational ground state\footnote{This can be seen by calculating the Boltzmann factor $\E^{-h\nu_2/k_B T}$, which equals 0.036 given $T=288$~K. This also indicates that neglecting stimulated emission is a good approximation in this case.}, so the fundamental band vibrational occupancy factor can be approximated as 1 for our purposes. For the rotational state, however, occupancy up to $J$ values of around 50 is significant at Earth-like temperatures. We take this into account by writing
\begin{equation}
\mathcal O_{mn}(T) = (2J+1)\E^{-J(J+1)hB/k_BT}/q_r
\end{equation}
with the rotational partition function approximated as $q_r \approx k_BT/2hB$ \citep[p. 172]{bernath2020spectra}. The $(2J+1)$ term in this expression accounts for degeneracy: i.e., for a given $J$ value, there are $2J+1$ states with the same energy.

The final step is to use line intensities to calculate the absorption spectrum itself. This is done by noting that the absorption cross-section for a single line, in units of m$^2$~molecule$^{-1}$, is simply 
\begin{equation}
\sigma_{mn} = S_{mn}f(\nu-\nu_{mn}).
\end{equation}
and the total absorption cross-section for the entire band is the sum of $\sigma_{mn}$ over all transitions,
\begin{equation}
\sigma(\nu) = \sum_{mn}\sigma_{mn}.
\end{equation}

Now we have all the pieces required to begin building absorption spectra. Using the line locations from Section~\ref{sec:linepos}, lineshape and width definitions from Section~\ref{sec:linewidth}, and line intensity formulae from this section, we can plot the P, Q and R branches of the $\nu_2$ fundamental band (Fig.~\ref{fig:analytic_spec}). Here, we have included lines up to $J=100$, on the basis that lines from higher rotational number transitions are so weak that they contribute little further to absorption. Comparison with Fig.~\ref{fig:HITRAN_spec} shows that line peaks are a little higher than in the HITRAN spectrum, mostly because we are somewhat underestimating the line broadening coefficient $\gamma$ via \eqref{eq:gamma_eqn} and \eqref{eq:gamma0_eqn}. Otherwise, the overall form of the band center is reproduced fairly accurately. At the sides of the band, however, it is clear that our approach fails completely. Correctly incorporating these sidebands will be the task of the next section.

\section{Fermi Resonance}\label{sec:Fermi}

\begin{figure}[h!]
	\begin{center}
		\includegraphics[width=3.0in]{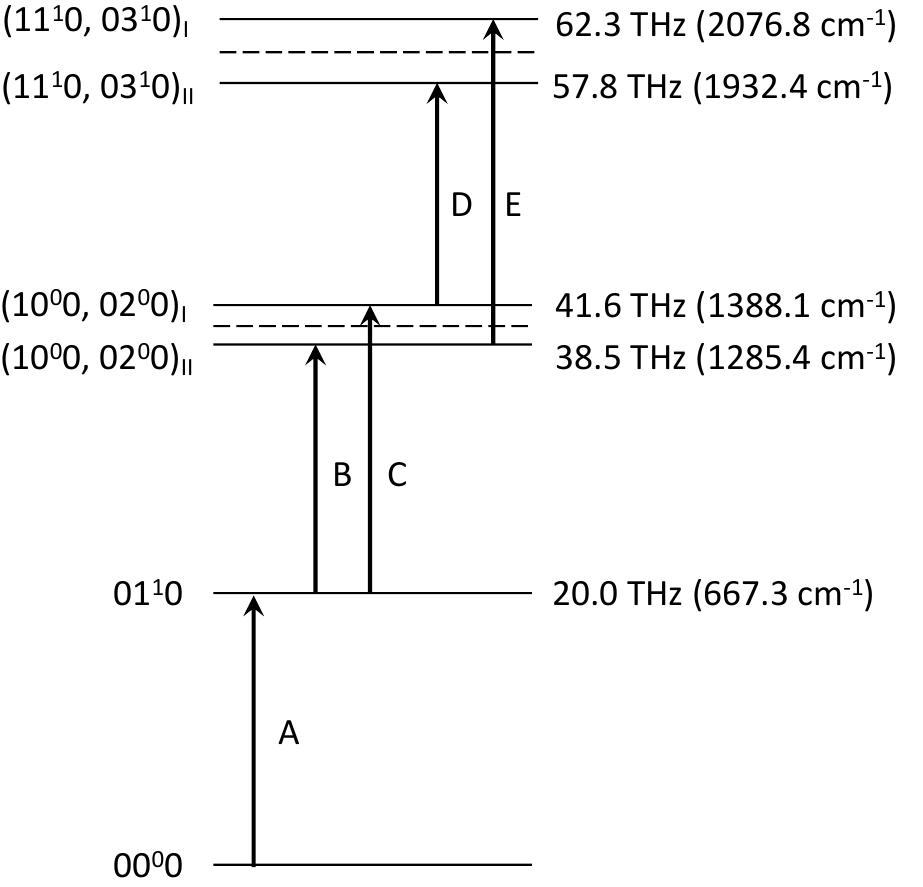}
	\end{center}
	\caption{The key vibrational energy levels and transitions of the \ce{CO2} molecule that are most important to radiative forcing and climate change. Many levels and transitions have been neglected for simplicity. Comparing with the notation of \cite{shine2023radiative}, our A $\equiv$ Fu; B, C $\equiv$ F1; D, E $\equiv$ FF, and H1, H2 and F2 are neglected. Numerical values for the energy levels were obtained from \cite{oberly1968bands}.}
\label{fig:energy_levels}
\end{figure}

The multiple sidebands present in the spectrum in Fig.~\ref{fig:HITRAN_spec} arise due to vibrational transitions between higher energy states than the $|00^00\rangle$ fundamental (Fig.~\ref{fig:energy_levels}). In a perfect simple harmonic oscillator, the spacing of vibrational energy levels is uniform, and all transitions occur at the same frequency. This is clearly not the case for the \ce{CO2} $\nu_2$ band. The reason for the difference is \emph{Fermi resonance} \citep{fermi1931ramaneffekt,adel1933infrared}. Fermi resonance is far better known in quantum spectroscopy than in climate physics, but it is key to understanding \ce{CO2} radiative forcing. 
{ Recently, \cite{shine2023radiative} extracted individual bands from the HITRAN database for the purpose of numerical calculations of the radiative forcing and found that Fermi resonance contributes approximately half of the total forcing magnitude. Here we begin by discussing the physical nature of Fermi resonance, and then show how incorporating it in our derivation allows us to write down a quantum analytic formula for \ce{CO2} radiative forcing.}

Fermi resonance occurs in \ce{CO2} because $\nu_1 \approx 2\nu_2$: the symmetric stretch frequency happens to be very close in value to double the bending frequency. As a result, nonlinear interactions between the two modes shift the energy levels of states $|10^00\rangle$ and $|02^00\rangle$ and cause their wavefunctions to mix (Fig.~\ref{fig:energy_levels}). The best way to get an intuitive understanding of Fermi resonance is by analogy with the classical coupled pendulum experiment. This analogy is noted in passing in \cite{herzberg1945molecular}, but we explore it in more detail here. 

In the coupled pendulum experiment, two pendulums of almost equal natural frequencies exchange energy with each other via some nonlinear coupling (most commonly, torsion of the string to which they are attached). Without this nonlinear interaction, a Fourier transform of their motion would yield a single peak. However, when the pendulums are coupled, the same transform yields two peaks that are shifted from the original central value by an amount that depends on the strength of the interaction.

\begin{figure}[h!]
	\begin{center}
		\includegraphics[width=6.0in]{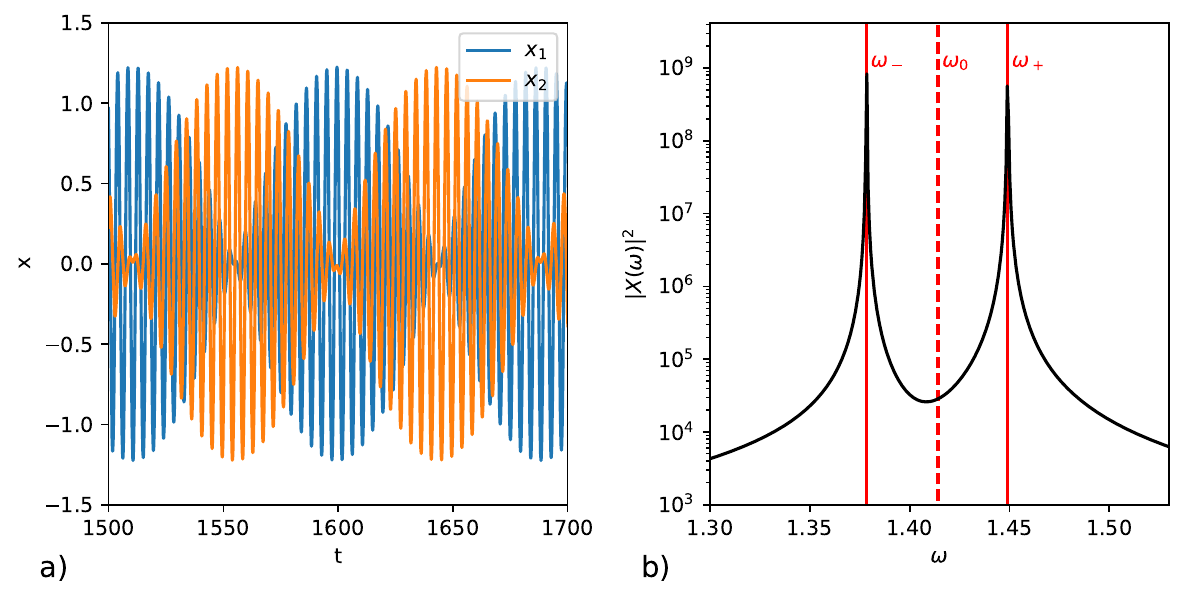}
	\end{center}
	\caption{Classical analogy to Fermi resonance. a) Numerical solution of \eqref{eq:coupled1} and \eqref{eq:coupled2} and b) the resulting power spectrum of $x_1$, with the analytical values of $\omega_0$, $\omega_-$ and  $\omega_+$ also shown. The simulation was run from $t=0$ to $t=10^4$ and solved using the 5(4) order Runge-Kutta method. In the analogy, $\omega_0$ is equivalent to $\nu_2$ (times a factor of 2$\pi$), and $\omega_-$ and $\omega_+$ are equivalent to the centers of the first Fermi sidebands.}
\label{fig:coupled_classical_oscillator}
\end{figure}

Mathematically, we can understand this process by first writing the classical Hamiltonian for a simple harmonic oscillator in isolation:
\begin{equation}
H(x,p) = \frac 12 \frac{ p^2}m + \frac 12 k x^2.
\end{equation} 
Here $p$ is momentum, $m$ is mass, $k$ is a force constant and $x$ is position. For two oscillators with a nonlinear coupling constant $\alpha$, the Hamiltonian is  
\begin{equation}
H(x_1,p_1,x_2,p_2) = \frac 12 \frac{ p_1^2}m +\frac 12 \frac{ p_2^2}m + \frac 12 k x_1^2 + \frac 12 k x_2^2 + \alpha x_1x_2
\end{equation} 
Using $d \mathbf p / dt = -\partial H/\partial \mathbf x$ and $d\mathbf x/dt = +\partial H/\partial \mathbf p$, where $\mathbf x = (x_1,x_2)$ and $\mathbf p = (p_1,p_2)$, we can derive equations of motion for each oscillator in the coupled case:
\begin{eqnarray}
\frac{d^2 x_1}{dt^2} &=&-\omega_0^2 x_1 - \tilde \omega^2 x_2, \label{eq:coupled1}\\
\frac{d^2 x_2}{dt^2} &=&-\omega_0^2 x_2 - \tilde\omega^2 x_1. \label{eq:coupled2}
\end{eqnarray} 
Here $\omega_0 = \sqrt{k/m}$ and $\tilde \omega = \sqrt{\alpha/m}$. The coupling term $\tilde \omega$ causes energy to be repeatedly exchanged between the first and second oscillators (Fig.~\ref{fig:coupled_classical_oscillator}). Taking the sum and difference of \eqref{eq:coupled1} and \eqref{eq:coupled2} and writing $ y_1 = x_1 + x_2$, $ y_2 = x_1 - x_2$, we have 
\begin{equation}
\frac{d^2  y_1}{dt^2}  =- \omega_+^2  y_1
\end{equation} 
\begin{equation}
\frac{d^2  y_2}{dt^2} =- \omega_-^2  y_2.
\end{equation} 
where $\omega_-^2 = \omega_0^2-\tilde\omega^2$ and $\omega_+^2 = \omega_0^2+\tilde\omega^2$. With this change of variables, it can be seen that the nonlinear coupling gives rise to two new eigenfunctions whose frequencies are shifted away from the unperturbed frequency $\omega_0$ by an amount dependent on the coupling constant $\alpha$. This is directly analogous to the way in which nonlinear coupling between the symmetric stretch and bending modes of \ce{CO2} give rise to the Fermi bands. Figure~\ref{fig:coupled_classical_oscillator} shows non-dimensional numerical solutions of \eqref{eq:coupled1} and \eqref{eq:coupled2} for $k=2$, $m=1$ and $\alpha = 0.1$, alongside the resulting power spectrum for $x_1$. 

To represent Fermi resonance in a quantum framework, a degenerate first-order perturbation analysis can be used \citep{dennison1940infrared}. For this, the molecular vibrational potential energy $U$ discussed earlier is expressed in terms of dimensionless mass-weighted normal coordinates $\zeta$, $\rho$ and $\sigma$ (see Fig.~\ref{fig:mode_schematic}) and expanded as $U = U^0 + \epsilon U^1 + ...$, where $\epsilon$ is a perturbation parameter. This results in the expressions
\begin{eqnarray}
U^0 &=& \frac 12 h(\nu_1\sigma^2 + \nu_2\rho^2 + \nu_3\zeta^2), \label{eq:U0}\\
\epsilon U^1 &=& \frac 12 h(a\sigma^3 + b\sigma \rho^2 + c' \sigma \zeta^2).  \label{eq:U1}
\end{eqnarray}
Here $a$, $b$ and $c'$ have units of frequency. \eqref{eq:U0} allows for simple harmonic oscillation of the \ce{CO2} molecule at its three normal frequencies $\nu_1$, $\nu_2$ and $\nu_3$, and nothing else. \eqref{eq:U1} contains anharmonic terms, of which the second, $b\sigma \rho^2$, describes interaction between the bending and symmetric stretch modes. Because the potential energy must remain the same if the coordinates are reflected across a plane of symmetry of the molecule, all other cubic terms ($\rho^3$, $\zeta^3$ etc.) in $U^1$ are forbidden for a linear molecule like \ce{CO2}.

With $U$ defined, the Hamiltonian of the system is written in the form $H = H^0+\epsilon H'$, and first-order degenerate perturbation theory {\citep[e.g., ][]{robinett2006quantum} can be employed to derive a formula for the perturbed energy levels and wavefunctions. 
For clarity, we focus on the nearly degenerate states $|10^00\rangle$ and $|02^00\rangle$, which we assign wavefunctions $\psi_m$ and $\psi_n$. Writing $\psi = a_m\psi_m + a_n\psi_n$, we can express the time-independent Schr\"odinger equation in the matrix form $\mathbf H \mathbf a = E\mathbf a$, where the individual elements of $\mathbf H$ are $\langle \psi_i | H|\psi_j  \rangle = \int  {\psi_i}^* H \psi_j d^3\mathbf x$. With the expansion $H = H^0+\epsilon H'$, we then have 
\begin{equation}
\begin{pmatrix}
E_m^0 + \epsilon W_{mm} & \epsilon W_{mn} \\
\epsilon W_{nm} & E_n^0 + \epsilon W_{nn}
\end{pmatrix}
\begin{pmatrix}
a_m \\
a_n
\end{pmatrix}
= E
\begin{pmatrix}
a_m  \\
a_n
\end{pmatrix}
\end{equation} 
where $W_{nm} = \langle \psi_n^0|H'| \psi_m^{0}\rangle$ and $E^0_m$ and $E^0_n$ are the unperturbed energies. Because of the symmetry of the system, $W_{mm} = W_{nn} = 0$ and $W_{nm} = W_{mn}^*$. Hence
\begin{equation}
\begin{pmatrix}
E_m^0 - E  & \epsilon W_{mn} \\
\epsilon W_{mn}^* & E_n^0 - E 
\end{pmatrix}
\begin{pmatrix}
a_m \\
a_n
\end{pmatrix}
= 0.
\end{equation} 
Taking the determinant of the matrix to solve the eigenvalue equation, we have 
\begin{equation}
(E_m^0  - E)(E_n^0  - E) - \epsilon^2 W_{mn}^*W_{mn}=0.
\end{equation} 
Writing $(E_m^0+E_n^0)/2 = \overline E^0$, setting $\epsilon=1$ and solving for $E$ yields
\begin{equation}
E_\pm= \overline E^0\pm\frac 12 \sqrt{ (E_m^0-E_n^0)^2 + 4 |W_{mn}|^2}.
\end{equation} 
Explicit evaluation of the perturbation matrix element for states $|m\rangle=|10^00\rangle$ and $|n\rangle=|02^00\rangle$ \citep{adel1933infrared,herzberg1945molecular} yields the result that $|W_{10^00;02^00}| = hb/\sqrt{2}$. Writing the difference between state frequencies in the absence of resonance interaction as $\Delta_0=(E_m^0-E_n^0)/h$, we have the result that resonance makes two new states emerge with energy levels separated by $\Delta_F = \pm  \sqrt{\Delta_0^2+2b^2}$.} At the level of approximation we are using, the frequency difference $|\Delta_0| = 0.5$~THz (16.7~\invcm), and the strength of coupling between the two modes $b = 2.14$~THz (71.3~cm$^{-1}$), so  $\Delta_F = 3.07$~THz (102.3~cm$^{-1}$). Interestingly, these values of $|\Delta_0|$ and $b$ were originally first determined from the spectra of \ce{CO2} itself \citep[both absorption and Raman;][]{fermi1931ramaneffekt,dennison1940infrared}.

\begin{figure}[h!]
	\begin{center}
	\includegraphics[width=6.5in]{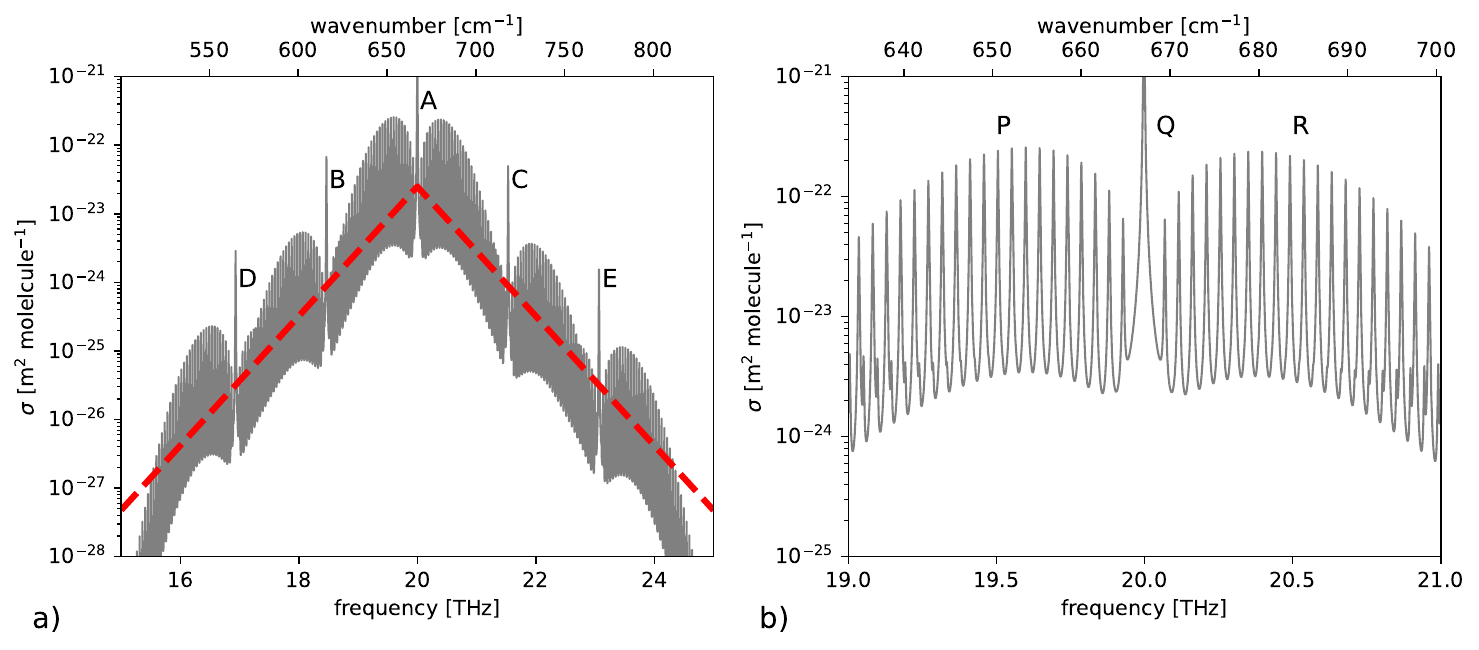}
	\end{center}
	\caption{Analytic \ce{CO2} $\nu_2$ band, including the fundamental vibrational transition and four additional Fermi sidebands. The red line in panel a) shows the approximate band shape predicted using \eqref{eq:band_struc_init} and \eqref{eq:band_struc}, given $T=T_s$. Labels in a) correspond to the transitions in Fig.~\ref{fig:energy_levels}. b) shows a close-up of the fundamental band. Pressure and temperature are the same as in Fig.~\ref{fig:HITRAN_spec}.}
\label{fig:analytic_spec_2}
\end{figure}

\section{\label{sec:radforce}Radiative Forcing}

Using the results of the previous two sections, we can now plot a revised synthetic spectrum for the \ce{CO2} $\nu_2$ band. The new Fermi resonance bands B and C (Fig.~\ref{fig:energy_levels}) are modeled just like the fundamental band, with the difference that they are shifted to lower/higher frequencies by an amount $|\Delta_F|/2$, and to lower line intensities by addition of a vibrational occupancy factor to $\mathcal O_{mn}(T)$. We take this to be the ratio of occupancy of the ground states for each transition, or $\E^{-h\nu_2/k_BT}$. We add the second set of Fermi bands D and E in a similar way, shifting them to lower/higher frequencies by a factor $|\Delta_F|$ and to lower lines intensities by $\E^{-2h\nu_2/k_BT}$. Recent analysis of radiative transfer model results shows that the fundamental band A contributes about 50\% of \ce{CO2} radiative forcing, while the additional Fermi bands B to E contribute most of the remaining 50\% \citep{shine2023radiative}. The contribution of other bands, such as the hot bands of the $\nu_2$ fundamental and the `second Fermi' bands of \cite{shine2023radiative}, are neglected here on the basis that they do not cause sufficiently large changes in absorption in the spectral regions where they are present. {Examples of such bands include the H1, H2 and F2 transitions mentioned in the caption to Fig.~\ref{fig:energy_levels}.}

 As can be seen from Fig.~\ref{fig:analytic_spec_2}, inclusion of all the labeled transitions in Fig.~\ref{fig:energy_levels} results in a far more realistic mid-infrared absorption band for \ce{CO2}. Reference to Fig.~\ref{fig:HITRAN_spec} shows that bands B and C are well captured by our approach. For the bands D and E, the limitations of our simple methodology becomes more apparent, as they are both about 0.6~THz (20~\invcm) too close to the fundamental band and too strong compared to HITRAN data. These differences arise due to our neglect of higher order anharmonic terms, and our assumption of a constant value for the Einstein-A coefficient. 
 
 Overall, however, inclusion of these four new Fermi bands results in a far more realistic mid-infrared absorption band for \ce{CO2} than was shown in Fig.~\ref{fig:analytic_spec}. It is worth emphasizing that despite the fact it reproduces the main P, Q and R branches visible in the HITRAN \ce{CO2} spectrum (Fig.~\ref{fig:HITRAN_spec}), the spectrum in Fig.~\ref{fig:analytic_spec} is vastly simpler. The total number of lines in this spectrum is 750, whereas in the HITRAN \ce{CO2} dataset over the 15--25~THz (500 to 833~\invcm) spectral range, around 100,000 lines are present if all isotopologues are included. Many of these additional lines are not visible in Fig.~\ref{fig:HITRAN_spec} because they are overlaid by the five strong bands incorporated in Fig.~\ref{fig:analytic_spec_2}.

We can also use the understanding developed in the last section to write down an analytic expression for the band structure coefficient $w$, and hence for \ce{CO2} radiative forcing itself. Clearly, $\nu_{cen} = \nu_2$, and $w$ must approximately equal half of $|\Delta_F|$ divided by the negative log of the occupancy factor:
\begin{equation}
w = \frac{|\Delta_F|}2 \frac{k_B T}{h \nu_2 }. \label{eq:band_struc}
\end{equation}
Given $T=250$~K and $\nu_2=20$~THz, $w = 0.40$~THz (13.3~cm$^{-1}$). The equivalent value obtained by fitting a full HITRAN spectrum\footnote{\cite{romps2022forcing}'s $b$, which has a value of $0.04$~cm, is equal to 1/2 times the inverse of our $w$ expressed in \invcm.} is remarkably close at 0.375~THz (12.5~\invcm) \citep{romps2022forcing}. 

Our value for $w$ yields $\alpha = 7.39$~\Wmsq~when substituted into \eqref{eq:alpha_defn}. This is somewhat larger than the 5.35~\Wmsq~value derived from comprehensive radiative transfer codes, {or the slightly higher effective radiative forcing values derived from Earth system models}\footnote{{Technical note: There are different definitions of radiative forcing \citep[e.g., ][]{masson2021climate}, which represents the perturbation to the planetary radiation budget due to a change in (in our case) of \ce{CO2}, in the absence of any surface temperature change. The simplest definition is the ``instantaneous radiative forcing'' in which only \ce{CO2} is changed. The ``stratosphere-adjusted radiative forcing'' allows stratospheric temperatures to adjust to the \ce{CO2} perturbation; this is a better predictor of surface temperature change, particularly when comparing the effect of changes in different constituents. IPCC's now preferred definition (``effective radiative forcing'') allows additional atmospheric adjustments (e.g., to tropospheric temperatures, humidity and clouds); it is an even better predictor, although detailed Earth-System Model calculations are needed to compute these adjustments. For the explanatory purposes of this paper, the differences between these definitions are of secondary importance.}}
 \citep{pinnock1995radiative,myhre1998new,masson2021climate}.  {The difference is primarily a result of the assumptions made in deriving \eqref{eq:alpha_defn}, and not our approach to calculating the \ce{CO2} spectrum, as demonstrated by the close correspondence of our $w$ value with that in \cite{romps2022forcing}. Refinement of the derivation leading up to \eqref{eq:alpha_defn} could be interesting to investigate in future.}

Finally, we can use the result just obtained to write an approximate expression for the radiative forcing from \ce{CO2} doubling as
\begin{equation}
\Delta F_{2\times} \approx \frac{\pi \ln (2) k_B T |\Delta_F|}{h \nu_2 }  [ \mathcal  B(\nu_{2},T_s) -  \mathcal  B(\nu_{2},T_t)] \approx 5.1~\mbox{\Wmsq}. \label{eq:QM_CC}
\end{equation}
Interestingly, almost all the background theory required to derive \eqref{eq:QM_CC} has existed since the first numerical \ce{CO2} radiative forcing calculations were performed in the 1960s. While this equation does not provide us with any truly new information, it does show us that the key driver of anthropogenic climate change can be expressed purely in terms of measured properties of Earth's atmosphere, and fundamental properties of the \ce{CO2} molecule, {without the need for any numerical calculations}. 

\section{\label{sec:climsens}Climate Sensitivity}

The preceding analysis shows that the value of \ce{CO2} radiative forcing can be understood in terms of the quantum properties of the \ce{CO2} molecule, with the influence of Fermi resonance on the band structure coefficient $w$ particularly critical. To tie this analysis back to \eqref{eq:ECS} in the introduction and link global warming to radiative forcing, the last thing we need is an expression for the climate feedback parameter $\lambda$. Analytic approaches to this problem have been discussed in detail in other recent works  \citep{ingram2010very,seeley2021h2o,jeevanjee2021simpson,koll2022analytical,stevens2022colorful,jeevanjee2023climate}, so we only provide a very brief summary here, in the interests of completeness. 

The simplest possible definition of $\lambda$ occurs on an airless planet, where all thermal emission comes from the surface. If the planet is also a blackbody, we have
\begin{equation}
\lambda = \frac{dF}{dT_s} = 4\sigma T_s^3 .\label{eq:lambda_simple}
\end{equation}
This yields $\lambda \approx 5.4$~\Wm2K for $T_s = 288$~K. This gives us an upper limit on $\lambda$ on Earth, because of course in many spectral regions the atmosphere is optically thick, and radiation to space occurs much higher, where air is colder and thermal emission is lower.

In-depth analysis in the papers cited above shows that under present-day conditions, emission in the mid-infrared \ce{H2O} window region is the dominant contributor to $\lambda$, so \eqref{eq:lambda_simple} can be approximated by 
\begin{equation}
\lambda = \frac{d}{dT_s}\int_{\nu_A}^{\nu_B} \pi \mathcal B(\nu,T_s) d\nu.  \label{eq:lambda_better}
\end{equation}
where $\nu_A$ and $\nu_B$ are the frequency limits of the window region. Using the Wien approximation, this integral can be written as 
\begin{equation}
\lambda \approx \frac{2\pi h^2\nu_w^4 \Delta \nu_w}{k_B c^2 T^2}  \E^{-h\nu_w/k_BT},
\end{equation}
where $\nu_w=(\nu_A+\nu_B)/2$ and $\Delta \nu_w = \nu_B - \nu_A$.

Writing\footnote{We leave the derivation of $\nu_A$ and $\nu_B$ from the quantum mechanical properties of the \ce{H2O} molecule to future work.} $\nu_A = 21$~THz ($700$~\invcm) and  $\nu_B = 36$~THz (1200~\invcm), we obtain $\lambda = 2.3$~\Wm2K, which is within the range for the clear-sky longwave feedback from observations and complex climate models \citep{andrews2012forcing,mckim2021joint,roemer2023direct}. Combining this estimate with the value for $\Delta F_{2\times}$ calculated in the last section yields $\Delta T_s = 2.2$~K. This is also reasonably close to values calculated using much more complicated models, despite the approximate nature of the calculation. As mentioned in the Introduction, {the range of} modern estimates for $\Delta T_s$ {extends to higher values} (2 to 5~K), in part because cloud feedbacks are predicted to increase warming in the future. 

\section{\label{sec:discuss}Discussion}

We have shown using mostly first-principles reasoning how the radiative forcing of \ce{CO2} emerges from the quantum mechanical properties of the \ce{CO2} molecule. This result has implications for our understanding of both contemporary global warming and the long-term evolution of Earth's climate. There are of course many things that our analysis misses out. Many spectroscopic details, including anharmonic interactions, line mixing, and additional weak bands have been neglected, as have overlap with other gaseous absorbers and the radiative effects of clouds. In common with many other 1D calculations, atmospheric vertical temperature structure has been treated crudely, and 3D dynamics is neglected entirely. Given all this, it is remarkable that our analysis and others like it still allows a reasonably accurate estimate of clear-sky radiative forcing and climate sensitivity. This outcome provides further evidence, if such evidence were needed, of the rock-solid foundation of the physics of global warming and climate change. 

{The work presented here should not be seen as a substitute for accurate numerical calculations, but instead a way to understand the fundamental physics that underpins them.} As our approach to spectroscopy in this paper is rather specific to the problem at hand, it is not immediately clear if there are many future applications for doing calculations from first principles in this way. Nonetheless, {it would be interesting to see if our approach could be extended to \ce{CO2} radiative forcing on other solar system planets. Mars is likely to be an easier case than Venus in the solar system, because of the importance of \ce{CO2} collision-induced absorption (CIA) in Venus's thick atmosphere, although in principle a simplified approach to CIA along the lines we have pursued here should also be possible}.  

Another interesting extension could be to use the analysis in Section~\ref{sec:method} to provide quick estimates of the warming potential of different greenhouse gases in a planetary context. {For exoplanet and paleoclimate applications, this might be a particularly useful way of increasing intuition and providing a reality check on the results of complex climate models.} It might also be interesting to investigate in more detail why resonances appear in certain molecules and bands but not in others. 

{Finally,} carbon dioxide has likely been a key greenhouse gas throughout Earth's history. Given this, the dependence of \ce{CO2} radiative forcing on the accidental resonance between $\nu_1$ and $\nu_2$ is particularly interesting. One can imagine that with minor differences in the quantum structure of \ce{CO2}, this resonance might be changed or inhibited, and the past and future evolution of our planet's climate would be very different. In this sense, the \ce{CO2} Fermi resonance may be seen as somewhat analogous to the nuclear resonances in astrophysics that give rise to the production of heavy elements in stellar interiors \citep{hoyle1954nuclear,livio1989anthropic}. 

\begin{acknowledgments}
We thank the organizers of the PCTS 2022 `From Spectroscopy to Climate: Radiative Constraints on the General Circulation' workshop for thought-provoking discussions that helped to motivate this article. {We also thank Nadir Jeevanjee and an anonymous reviewer for insightful comments that significantly improved the quality of this manuscript.} RW acknowledges additional discussions with Iouli Gordon on HITRAN unit conversion and related topics. RW and JS acknowledge funding from NSF awards AST-1847120 (CAREER) and AGS-2210757. Code to reproduce the plots in the paper is available open-source at \emph{https://github.com/wordsworthgroup/quantum\_global\_warming}. 
\end{acknowledgments}

\bibliography{main}
\bibliographystyle{aasjournal}

\newpage

\end{document}